\documentclass[12pt]{article}
\usepackage[cp1251]{inputenc}
\usepackage[russian,english]{babel}
\usepackage{amsfonts,amssymb}
\usepackage{graphicx}
\usepackage{amsmath, amssymb, graphics, setspace}
\usepackage{hyperref}

\begin{document}

\title{Geodesic motion of test particles in Korkina--Grigoryev metric}
\maketitle
\begin{center}
I.~Bormotova$^{1,2, *}$ \quad   E.~Kopteva$^{1,2}$  \\   
       $^1${\it Joint Institute for Nuclear Research, Dubna, Russia} \\
       $^2${\it Institute  of  Physics,  Faculty  of  Philosophy  and  Science,\\
Silesian  University  in  Opava, Czech Republic} \\
	$^*${e-mail: q\_Leex@mail.ru}\\
\end{center}

\begin{abstract}
We study the geodesic structure of the Korkina-Grigoryev spacetime. The corresponding metric is a generalization of the Schwarzschild geometry to the case involving a massless scalar field. We investigate the relation between the angular momentum of the test particle and the charge of the field, which determines the shape of the effective-potential curves. The ratio for angular momentum of the particle, the charge of the scalar field and the dimensionless spatial parameter is found, under which the finite motion of particles occurs. From the behavior of the potential curves the radii of both stable and unstable circular orbits around a black hole are found, as well as the corresponding energies of the test particles. The effective-potential curves for the Korkina-Grigoryev, the Schwarzschild and the Reissner-Nordstrom fields are compared. It is shown, that in the case of the Korkina-Grigoryev metric the stable orbits eventually vanish with increasing charge.
\end{abstract}

\section{Introduction}

Investigation of motion of test particles is a common method in general relativity to study the structure and properties of space-time near a gravitational object \cite{Lan}. In this paper we consider and analyze the motion of test particles in Korkina--Grigoryev  spacetime.

\section{Korkina--Grigoryev metric}

Metric for Korkina--Grigoryev (KG) gravitational field \cite{Kor}, induced by a nonlinear massless conformally-invariant scalar field, has the form:

\begin{equation}
\label{mKG}
{\rm d}s^{2} =\xi ^{3} \frac{r_{g} }{r} {\rm d}t^{2} -\frac{\xi }{(\xi ^{2} +b)^{2} } \cdot \frac{r}{r_{g} } {\rm d}r^{2} -r^{2} {\rm d}\sigma ^{2}, 
\end{equation}
\\
where $r=r_{g} (1+3b\xi +\xi ^{3} )$; $\xi $ is a dimensionless spatial variable; $b=3G^{2} r_{g}^{-2/3} $; \textit{G}  is a charge that characterizes the energy of the scalar field.

Potential of KG nonlinear scalar field is given by:

\begin{equation}
\phi =Gr_{g}^{1/3} \xi.
\end{equation}

Metric (\ref{mKG}) can be rewritten in a modified form

\begin{equation}
{\rm d}s^{2} =\left(1-\frac{r_{g} }{r} (1+3b\xi )\right){\rm d}t^{2} -\left(1-\frac{r_{g} }{r} \left(1+3b\xi -\frac{b^{2} }{\xi } \right)\right)^{{\rm -1}} {\rm d}r^{2} -r^{2} {\rm d}\sigma ^{2},
\end{equation}
\\
which is more convenient for analyzing and comparing with known metrics.

Using the mass function method \cite{Gla} we found the energy density in the case under consideration

\begin{equation}
\label{dens}
\varepsilon =\frac{G^{2} }{r_{g}^{2/3} } \cdot \frac{1}{\xi ^{2} r^{2} }, 
\end{equation}

From the expression (\ref{dens}) it follows that the influence of the scalar field decreases with the square of the distance.

\section{The effective potential and extreme functions}

Using the Hamilton--Jacobi method we find the expression for the effective potential in dimensionless variables:

\begin{equation}
\label{U}
U_{{\rm eff}} =\sqrt{\left(1-\frac{1+3b\xi }{1+3b\xi +\xi ^{3} } \right)\left(1+\frac{A}{(1+3b\xi +\xi ^{3} )^{2} } \right)}, 
\end{equation}
\\
where
\begin{align}
&U_{{\rm eff}} =\frac{U_{{\rm eff}} }{mc^{2} } \nonumber ,\\ 
&r=\frac{r}{r_{g} } =1+3b\xi +\xi ^{3} ,\\
 &b=3G^{2} r_{g}^{-2/3} ,\quad A=\frac{L^{2} }{m^{2} c^{2} r_{g}^{2} }.  \nonumber
\end{align}

Taking into account following conditions

\begin{equation}
U_{{\rm eff}}^{2} =E^{2} =1,\quad \frac{\partial U_{{\rm eff}} }{\partial r} =0 
\end{equation}
\\
for the effective potential (\ref{U}) we obtain the expressions for the radii of circular orbits and turning points:

\begin{align}
\label{ko}
& r_{\mathrm{min_{\pm}}}=\frac{A \pm \sqrt{A^2-3A-18A b \xi - 27 A b^2 \xi^2}}{1+3 b \xi}, \\
& r_{\mathrm{turn_{\pm}}}=\frac{A \pm \sqrt{A^2-4A-24A b \xi - 36 A b^2 \xi^2}}{2(1+3 b \xi)},
\end{align}
\\
where ``+''  corresponds to stable orbits and  ``\textbf{--}'' is for unstable ones.

From (\ref{U}) we found the condition for angular momentum \textit{A}, the charge of the field \textit{b}, and parameter $\xi $ under which the finite motion of particles will occur

\begin{equation}
A=\frac{(1+3b\xi +\xi ^{3} )^{2} (1+2b\xi )}{2\xi ^{3} -1}.
\end{equation}

\begin{figure}[h]
\begin{minipage}[h]{0.49\linewidth}
\center{\includegraphics[width=1\linewidth]{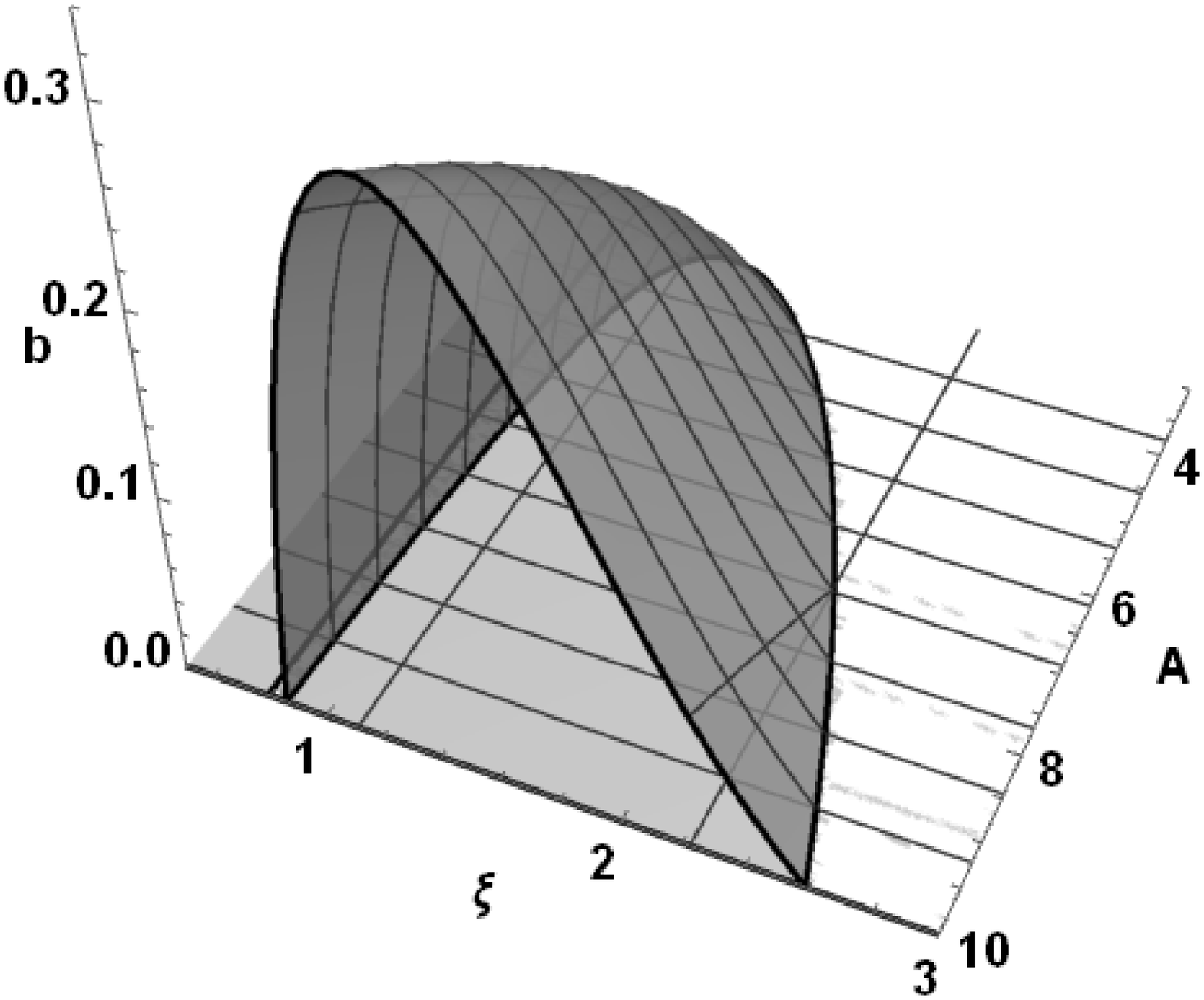} \\ (a)}
\end{minipage}
\hfill
\begin{minipage}[h]{0.49\linewidth}
\center{\includegraphics[width=1\linewidth]{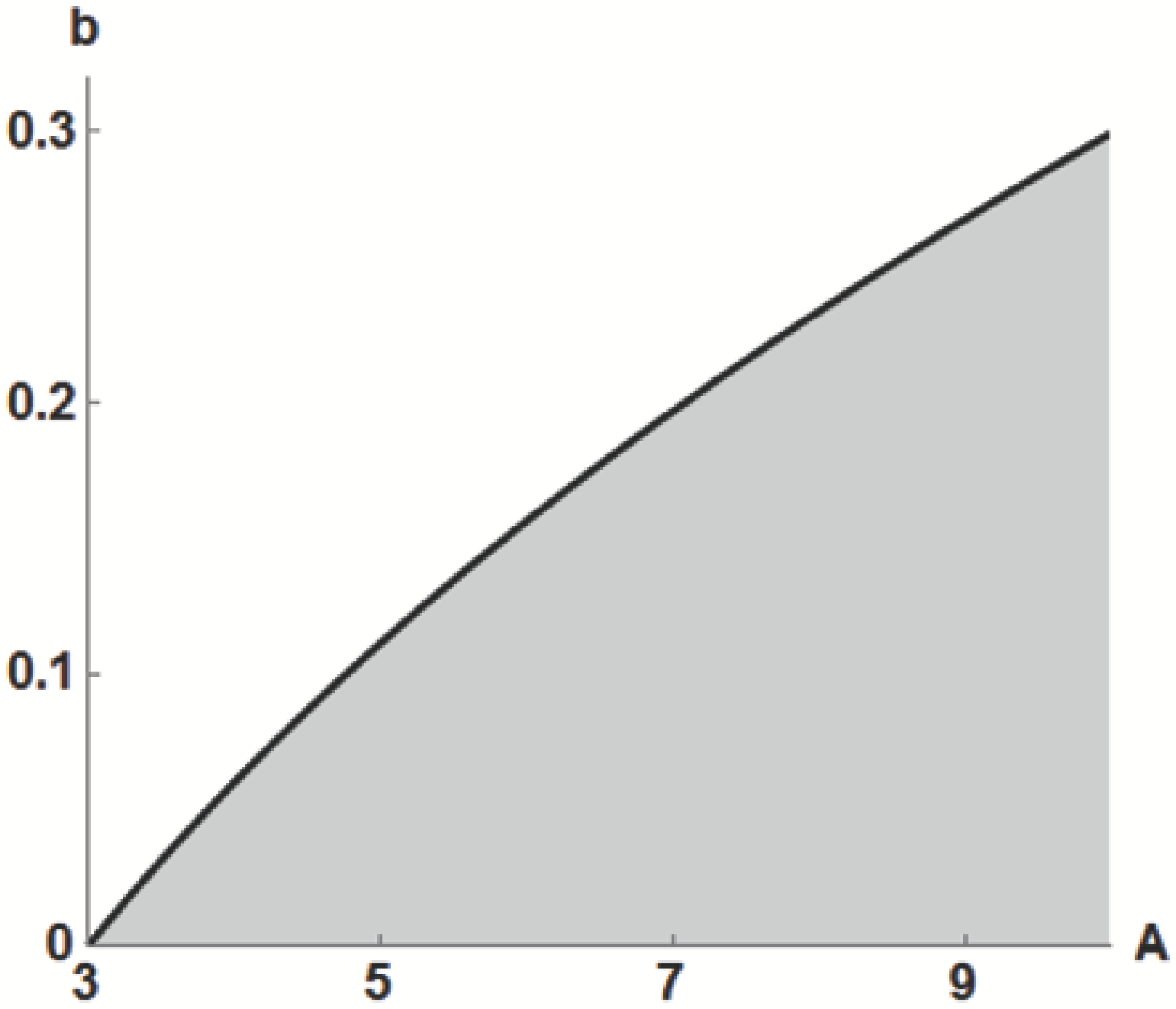} \\ (b)}
\end{minipage}
\caption{The region of values of the parameters \textit{A, b}, $\xi $ for finite motion of particles (a). The region of values of the parameters \textit{A, b} under fixed $\xi $ for finite motion of particles (b).}
\end{figure}

Fig.1a shows the region of values of parameters \textit{A}, \textit{b}, $\xi $ that allow finite motion of particles. Here the shaded surface corresponds to circular orbits, the area under the surface corresponds to elliptic orbits, $A\xi $-plane with \textit{b} = 0 corresponds to Schwarzschild-type black hole. 

Fig.1b shows the section of Fig.1a for values of the parameters \textit{A}, \textit{b} under fixed $\xi $, where the shaded area corresponds to the existence of elliptic orbits, and the bold boundary line corresponds to circular orbits, beyond this line a finite motion of particles is impossible.

From (\ref{U}) it follows that the parameter $\xi $ must always be a positive $\xi >\sqrt[{3}]{{1 \mathord{\left/{\vphantom{1 2}}\right.\kern-\nulldelimiterspace} 2} } $. Thus, the potential curves exist at

\begin{equation}
r>\sqrt{\frac{1}{2} } b+\frac{3}{2}.
\end{equation}

Fig. 2 shows a comparison of the potential curves for Korkina--Grigoryev field and known Schwarzschild and Reissner--Nordstr\"om fields \cite{Pug}. It can be seen from fig. 2 that under zero charges KG and Reissner--Nordstr\"om solutions transform into the Schwarzschild solution. If the charge of KG field is increasing, the potential curves will lose the extrema in the spatial domain, so that the finite motion of particles will become impossible, unlike the situation for the Reissner--Nordstr\"om solution. 

\begin{figure}[h!]
\centering
\includegraphics[width=0.5\linewidth]{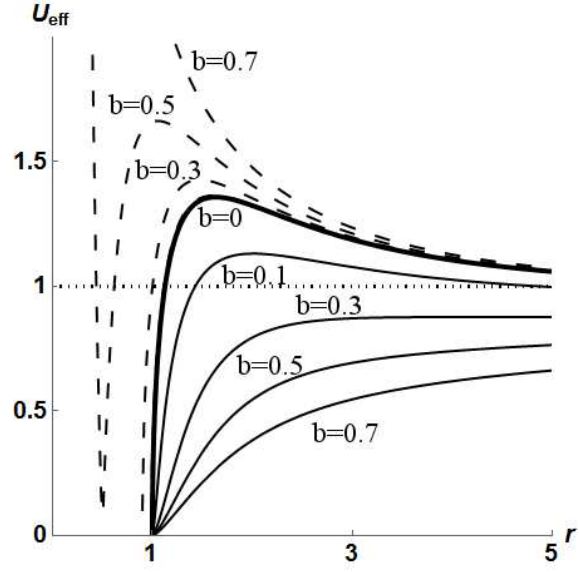}
\caption{The comparison of potential curves for KG field (solid lines), Schwarzschild field (bold line) and Reissner--Nordstr\"om field (dashed lines) for a different values of the correspondent charges.}
\end{figure}

\section*{Conclusions}

In this work the motion of test particles in Korkina--Grigoryev field is considered. The ratio between angular momentum \textit{A}, charge of field \textit{b} and the dimensionless parameter $\xi $ is found under which the finite motion of particles will occur. The lower limit of parameter $\xi >\sqrt[{3}]{{1 \mathord{\left/{\vphantom{1 2}}\right.\kern-\nulldelimiterspace} 2} } $ is obtained. It is shown that for large values of charge of KG field the orbits are disappear and the finite motion of particles is impossible. The comparison of the potential curves for Korkina--Grigoryev, Schwarzschild and Reissner--Nordstr\"om fields are considered.

\section*{Acknowledgments}

This work is supported by the Grant of the Plenipotentiary Representative of the Czech Republic in JINR under Contract No. 189 from 29/03/2016.

\end{document}